\documentclass[twocolumn]{autart}
\usepackage{epsfig}\usepackage{amsmath}

\def\e{\epsilon}

\begin{document}

\begin{frontmatter}

\title{Tracking Control for Multi-Agent Consensus with an Active Leader and Variable Topology}

\author[beijing]{Yiguang Hong}\ead{yghong@iss.ac.cn},    % Add the
\author[beijing]{Jiangping Hu}\ead{jphu@amss.ac.cn},               % e-mail address
\author[beijing,wenzhou]{Linxin Gao}\ead{lxgao@amss.ac.cn}  % (ead) as shown

\address[beijing]{Key Laboratory of Systems and Control, Institute of Systems
Science\\
Chinese Academy of Sciences, Beijing 100080, China }
\address[wenzhou]{Institute of Systems Science, Wenzhou University, Zhejiang, China}

\begin{keyword}
Multi-agent systems, consensus, state estimation, active leader.
\end{keyword}

\begin{abstract}
In this paper, we consider a multi-agent consensus problem with an
active leader and variable interconnection topology. The state of
the considered leader not only keeps changing but also may not be
measured.  To track such a leader, a neighbor-based local
controller together with a neighbor-based state-estimation rule is
given for each autonomous agent. Then we prove that, with the
proposed control scheme, each agent can follow the leader if the
(acceleration) input of the active leader is known, and the
tracking error is estimated if the input of the leader is unknown.
\end{abstract}

\end{frontmatter}

\section{Introduction}

In recent years, there has been an increasing research interest in
the control design of multi-agent systems. Many results have been
obtained with local rules applied to each agent in a considered
multi-agent system.  These neighbor rules for each agent are based
on the average of its own information and that of its neighbors or
its leader (Fax \& Murray, 2004; Jadbabaie, Lin, \& Morse, 2003;
Lin,  Broucke, \& Francis, 2004; Olfati-Saber \& Murray, 2004;
Savkin, 2004). For example, Jadbabaie {\it et al.} (2003)
demonstrated that a simple neighbor rule makes all agents
eventually move in the same direction despite the absence of
centralized coordination and each agent's set of neighbors
changing with time as the system evolves under a joint connection
condition. Also, with a similar technique, Lin {\it et al.} (2004)
studied three formation strategies for groups of mobile autonomous
agents. The stability analysis of multi-vehicle formations was
given with a Nyquist-type criterion in (Fax \& Murray, 2004).
Moreover, by a Lyapunov-based approach, Olfati-Saber {\it et al.}
(2004) solved the average-consensus problem with directed
interconnection graphs or time-delays.

In reality, some variables of the agents and/or the leader in a
multi-agent system may not be able to be measured. Fax {\it et
al.} (2004) raised this important issue regarding observer design
for multi-agent systems, and first tackled this problem. However,
many works remain to be done for the distributed observer design
of networks of multiple agents.

With this background, we consider a consensus problem with an
active leader with an underlying dynamics.  Here, some variables
(that is, the velocity and maybe the acceleration) of an active
leader cannot be measured, and each agent only gets the measured
information (that is, the position) of the leader once there is a
connection between them. In this paper, we propose an ``observer"
by inserting an integrator into the loop for each agent to
estimate the leader's velocity. To analyze the problem, a
Lyapunov-based approach is developed. With the proposed estimation
rule and a selected Lyapunov function, the leader-following
problem can be solved if the leader's input is known, while the
tracking error can also be analyzed if the input is unknown.

%This paper is organized as follows. In Section 2, our problem is
%formulated along with related preliminaries.  Then the main
%results are presented in Section 3. A local control scheme is
%given to handle the leader-following problems with an active
%leader and variable interconnection topology.  Based on state
%estimation, the convergence analysis is given with or without
%knowing the acceleration input of the leader. Finally, concluding
%remarks are made in Section 4.

\section{Problem Formulation}

To solve coordination problems, graph theory is helpful. An
undirected graph $\mathcal{G}$ on vertex set
$\mathcal{V}=\{1,2,\cdots,n\}$ contains $\mathcal{V}$ and a set of
unordered pairs $\mathcal{E}=\{(i,j):i,j\in \mathcal{V}\}$, which
are called $\mathcal{G}$'s edges. If there is an edge between two
vertices, the two vertices are called adjacent. A graph is simple
if it has no self-loops or repeated edges. If there is a path
between any two vertices of a graph $\mathcal{G}$, then
$\mathcal{G}$ is connected, otherwise disconnected. A subgraph
${\mathcal X}$ of $\mathcal{G}$ is an induced subgraph if two
vertices of ${\mathcal V(X)}$ are adjacent in ${\mathcal X}$ if
and only if they are adjacent in $\mathcal{G}$. An induced
subgraph ${\mathcal X}$ of $\mathcal{G}$ that is maximal, subject
to being connected, is called a component of $\mathcal{G}$.

Here we consider a system consisting of $n$ agents and a leader.
In the sequel, the state of agent $i$ is denoted by $x_i$ for
$i=1,...,n$. With regarding the $n$ agents as the vertices in
$\mathcal{V}$, the relationships between $n$ agents can be
conveniently described by a simple and undirected graph ${\mathcal
G}$, which is defined so that $(i,j)$ defines one of the graph's
edges in case agents $i$ and $j$ are neighbors.  $N_{i}(t)$
denotes the set of labels of those agents which are neighbors of
agent $i\; (i=1,...,n)$ at time $t$.  The weighted adjacency
matrix of ${\mathcal G}$ is denoted by $A=[a_{ij}]\in R^{n\times
n}$, where $a_{ii}=0$ and $a_{ij}=a_{ji}\geq 0$ ($a_{ij}>0$ if
there is an edge between agent $i$ and agent $j$).  Its degree
matrix $D=diag\{ d_1,...,d_n\}\in R^{n\times n}$ is a diagonal
matrix, where diagonal elements $d_i=\sum_{j=1}^n a_{ij}$ for
$i=1,...,n$. Then the Laplacian of the weighted graph is defined
as
\begin{equation}
L=D-A,
\end{equation}
which is symmetric.  In what follows, we mainly concern a graph
$\bar{\mathcal G}$ associated with the system consisting of $n$
agents and one leader.  In fact, $\bar{\mathcal G}$ contains $n$
agents (related to graph ${\mathcal G}$) and the leader with
directed edges from some agents to the leader. By ``the graph,
$\bar{\mathcal G}$, of this system is connected", we mean that at
least one agent in each component of ${\mathcal G}$ is connected
to the leader.

For the multi-agent system under consideration, the relationships
between neighbors (and the interconnection topology) change over
time. Suppose that there is an infinite sequence of bounded,
non-overlapping, contiguous time-intervals $[t_{i},t_{i+1}), \;
i=0,1,\cdots$, starting at $t_{0}=0$.

Denote ${\mathcal{S}}=\{\bar {\mathcal G}_1, \bar {\mathcal
G}_2,\cdots,\bar {\mathcal G}_N\}$ as a set of the graphs with all
possible topologies, which includes all possible interconnection
graphs (involving $n$ agents and a leader), and denote
$\mathcal{P}=\{1,2,\cdots, N\}$ as its index set.
%Obviously, ${\mathcal{S}}$ has finite elements.

To describe the variable interconnection topology, we define a
switching signal $\sigma: [0,\infty)\rightarrow \mathcal{P}$,
which is piecewise-constant. Therefore, $N_i$ and the connection
weight $a_{ij}\; (i=1,...,n, j=1,...,n)$ are time-varying, and
moreover, Laplacian $L_{p}\; (p\in {\mathcal P})$ associated with
the switching interconnection graph is also time-varying (switched
at $t_i,\; i=0,1,\cdots$), though it is a time-invariant matrix in
any interval $[t_{i},t_{i+1})$. In our problem, we assume that
there are fixed positive constants $\alpha_{ij}\; (i=1,...,n;
j=1,...,n)$ such that
\begin{equation}
\label{alpha} a_{ij}(t)=\begin{cases} \alpha_{ij}=\alpha_{ji},&\mbox{if agents $i$ and $j$}\\
&\mbox{\qquad are connected at $t$} \\
0,& \mbox{otherwise}\end{cases}
\end{equation}
Meanwhile, the connection weight between agent $i$ and the leader,
denoted by $b_i$, is time-varying, too.  We assume that there are
fixed positive constants $\beta_i\; (i=1,...,n)$ such that
\begin{equation}
\label{beta} b_i(t)=\begin{cases} \beta_i &\mbox{if agent $i$ is
connected to the leader at $t$} \\ 0 & \mbox{otherwise}\end{cases}
\end{equation}

The next lemma was given in Horn and Johnson (1985), to check the
positive definiteness of a matrix.

\begin{lem}
\label{lem1} Suppose that a symmetric matrix is partitioned as
$$
E=\begin{pmatrix}E_1&E_2 \\ E_2^{T}&E_3 \end{pmatrix}
$$
where $E_1$ and $E_3$ are square. $E$ is positive definite if and
only if both $E_1$ and $E_3- E_2^{T}E_1^{-1}E_2$ are positive
definite.
\end{lem}

The following result is well-known in algebraic graph theory
(Godsil \& Royle, 2001) and establishes a direct relationship
between the graph connectivity and its Lapalcian.

\begin{lem}
\label{lem2} Let $\mathcal{G}$ be a graph on $n$ vertices with
Laplacian $L$. Denote the eigenvalues of $L$ by
$\lambda_{1}(L),\cdots, \lambda_{n}(L)$ satisfying
$\lambda_{1}(L)\leq\cdots\leq\lambda_{n}(L)$. Then
$\lambda_{1}(L)=0$ and $\textbf{1}=[1,1,\cdots,1]^{T}\in R^n$ is
its eigenvector. Moreover, if ${\mathcal G}$ is connected,
$\lambda_2>0$.
\end{lem}

In this paper, all the considered agents move in a plane:
\begin{equation}
\label{modela} \dot x_i=u_i\in R^2,\quad i=1,...,n,
\end{equation}
where $u_i$ is the control input. The leader of this considered
multi-agent system is active; that is, its state variables keep
changing.  Its underlying dynamics can be expressed as follows:
\begin{equation}
\label{model0} \begin{cases}\dot{x}_0=v_0\\ \dot v_0=a(t)=
a_0(t)+\delta(t)\\ y=x_0\end{cases}\quad x_0,\, v_0,\, \delta\in
R^2
\end{equation}
where $y(t)=x_0(t)$ is the measured output and $a(t)$ is the
(acceleration) input.  Note that (\ref{model0}) is completely
different from the agent dynamics (\ref{modela}).  In other words,
the agents will track a leader with a different dynamics.

In our problem formulation, the input $a(t)$ may not be completely
known.  We assume that $a_0(t)$ is known and $\delta(t)$ is
unknown but bounded with a given upper bound $\bar\delta$ (that
is, $||\delta(t)||\leq \bar \delta$).  The input $a(t)$ is known
if and only if $\bar\delta=0$.  On the other hand, $y=x_0$ is the
only variable that can be obtained directly by the agents when
they are connected to the leader.  Our aim here is to propose a
decentralized control scheme for each agent to follow the leader
(i.e., $x_i \to x_0$).

Since $v_0(t)$ cannot be measured even when the agents are
connected to the leader, its value cannot be used in the control
design. Instead, we have to estimate $v_0$ during the evolution.
Note that, each agent has to estimate $v_0$ only by the
information obtained from its neighbors in a decentralized way.
The estimate of $v_0(t)$ by agent $i$ is denoted by $v_i(t)$
($i=1,...,n$). Therefore, for each agent, the local control scheme
consists of two parts:
\begin{itemize}
\item a neighbor-based feedback law:
\begin{equation}
\label{con} \begin{split} u_i=&-k[\sum_{j\in N_i(t)}a_{ij}(t)(x_i-
x_j)+b_i(t)(x_i-x_0)]\\
&+v_i,\quad k>0, \; i=1,\cdots,n,
\end{split}
\end{equation}
where $N_i$ is the set consisting of agent $i$'s neighbor agents;

\item a dynamic neighbor-based system to estimate $v_0$
\begin{equation}
\label{update1}
\begin{split}
 \dot v_i=&a_0-\gamma k[\sum_{j\in
N_i(t)}a_{ij}(t)(x_i- x_j)+b_i(t)\cdot\\
&(x_i-x_0)],\quad i=1,\cdots,n,
\end{split}
\end{equation}
for some positive constant $\gamma<1$.  In fact, (\ref{update1}),
can be viewed as an ``observer" in some sense.
\end{itemize}
%auxiliary

Note that $u_i$ in (\ref{con}) is a local controller of agent $i$,
which only depends on the information from its neighbors, and, in
fact, when $v_0=0, a=0$, the proposed control law (\ref{con}) is
consistent with the one given in Olfati-Saber and Murray (2004).
In addition, with the neighbor-based estimation rule in a form of
observer (\ref{update1}) to estimate the leader's velocity, each
agent relies only on the locally available information at every
moment. In other words, each agent cannot ``observe" or
``estimate" the leader directly based on the measured information
of the leader if it is not connected to the leader. In fact, it
has to collect the information of the leader in a distributed way
from its neighbor agents.

Take
$$
x=\begin{pmatrix} x_1\\ \vdots \\ x_n\end{pmatrix}, \quad v=\begin{pmatrix} v_1\\
\vdots \\ v_n\end{pmatrix},\quad u=\begin{pmatrix} u_1\\
\vdots \\ u_n\end{pmatrix}.
$$
Regarding the switching interconnection graphs, the closed-loop
system can be expressed as:
\begin{equation}
\label{model1}
\begin{cases}
\dot{x}=u=-k(L_{\sigma}+B_{\sigma})\otimes I_2 x+k
B_{\sigma}\textbf{1}\otimes x_{0}+v
\\
\dot v =\textbf{1}\otimes a_{0}-\gamma k(L_\sigma+B_\sigma)\otimes
I_2 x+\gamma k (B_{\sigma}\textbf{1})\otimes x_{0}
\end{cases}
\end{equation}
where $I_l\in R^{l\times l}$ (for any positive integer $l$) is the
identity matrix and $\otimes$ denotes the Kronecker product,
$\sigma: [0,\infty)\to \mathcal{P}=\{1,2,\cdots,N\}$ is a
piecewise constant switching signal with successive switching
times, $B_{\sigma}$ is an $n\times n$ diagonal matrix whose $i$th
diagonal element is $b_i(t)$ at time $t$, $L_{\sigma}$ is the
Laplacian for the $n$ agents.  Note that, even in the case when
the interconnection graph is connected, $b_i(t)$ may be always 0
for some $i$, and therefore, $B_{\sigma}$ may not be of full rank.

Denote $\bar x=x-\textbf{1}\otimes x_{0}$ and $\bar v=v-
\textbf{1}\otimes v_{0}$. Because $-k (L_\sigma+B_\sigma)\otimes
I_2 x+k B_{\sigma}\textbf{1}\otimes x_{0} =-k
(L_\sigma+B_\sigma)\otimes I_2\bar x$ (invoking Lemma \ref{lem2}),
we can obtain an error dynamics of (\ref{model1}) as follows:
\begin{equation}
\label{model2} \dot{\epsilon}=F_{\sigma}\epsilon+g,\quad
g=\begin{pmatrix} 0\\ -{\bf 1}\otimes \delta
\end{pmatrix}
\end{equation}
where
$$
\epsilon=\left(\begin{array}{c} \bar x\\
\bar v\end{array}\right),\quad F_{\sigma}
=\begin{pmatrix} -k(L_\sigma+B_\sigma)&I_{n}\\
-\gamma k(L_\sigma+B_\sigma)&0
\end{pmatrix}\otimes I_2.
$$

\section{Main Results}

In this section, we investigate the consensus problem of
multi-agent system (\ref{model1}), or the convergence analysis of
system (\ref{model2}). If the information of the input $a(t)$ can
be used in local control design, we can prove that all the agents
can follow the leader, though the leader keeps changing. If not,
we can also get some estimation of the tracking error.  We first
assume that the interconnection graph $\bar{\mathcal G}$ is always
connected, though the interconnection topology keeps changing; and
then we consider an extended case.

As mentioned above, $\bar{\mathcal G}$ is connected if at least
one agent in each of its component is connected with the leader.
To be specific, if there are $m\geq 1$ components, then the
Laplacian $L_{p}$ (for any $p\in {\mathcal P}$) of the graph
associated with $n$ agents have $m$ zero eigenvalues. For
simplicity, we can rearrange the indices of $n$ agents such that
$L_{p}$ can be rewritten as a block diagonal matrix:
$$
L_{p}=\left(
\begin{array}{cccc}
L_p^{1}&&&\\
&L_p^{2}&&\\
&&\ddots&\\
&&&L_p^{m}
\end{array}
\right)
$$
where each block matrix $L_p^i$ is also a Laplacian of the
corresponding component.  For convenience, denote
$M_{p}=L_{p}+B_{p}$, where $L_p$ is the weighted Laplacian and
$B_p\; (p\in {\mathcal P})$ is the diagonal matrix as defined in
Section 2.  The next lemma is given for $M_p$.

\begin{lem}
\label{lem3} If graph $\bar {\mathcal G}_p$ is connected, then the
symmetric matrix $M_p$ associated with $\bar {\mathcal G}_p$ is
positive definite.
\end{lem}

Proof: We only need to prove the case when $m=1$. Let
$\lambda_1,\cdots,\lambda_{n}$ be the eigenvalues of Laplacian
$L_p$ in the increasing order. From Lemma \ref{lem2},
$\lambda_{1}=0$ and $\lambda_{i}>0,i\geq 2$. Denote $n$
eigenvectors of $L_p$ by $\zeta_{i},\; i=1,...,n$, with
$\zeta_1=\textbf{1}$, an eigenvector of $L_p$ corresponding to
$\lambda_1=0$. Then any nonzero vector $z\in R^{n}$ can be
expressed by $z=\sum_{i=1}^{n}c_i\zeta_{i}$ for some constants
$c_i,i=1,2,\cdots,n$. Moreover, $B_p\neq 0$ since there is at
least one agent connected to the leader.  Without loss of
generality, we assume $b_j> 0$ for some $j$, and it is obvious
$\zeta_1^TB_p\zeta_1\geq b_j$.  Therefore, in either the case when
$c_2=...=c_n=0$ (so $c_1\neq 0$) or the case when $c_i\neq 0$ for
some $i\geq 2$, we always have
\begin{equation*}
\label{esti1} z^{T}M_pz=z^{T}L_pz+z^{T}B_pz \geq
\sum_{i=2}^{n}\lambda_{i}c_{i}^{2}\zeta_i^T\zeta_i +z^{T}B_pz>0
\end{equation*}
for $z\neq 0,$ which implies the conclusion. \hfill\rule{4pt}{8pt}

Based on Lemma \ref{lem3} and the fact that the set ${\mathcal P}$
is finite,
\begin{equation}
\label{barlamb} \begin{split} \bar
\lambda=&\min\{\mbox{eigenvalues of}\; M_p\in R^{n\times n},\;
\forall \mbox{$\bar {\mathcal G}_p$
is} \\
& \mbox{connected}\} >0,
\end{split}
\end{equation}
is fixed and depends directly on the constants $\alpha_{ij}$ and
$\beta_i$ for $i=1,...,n,\; j=1,...,n$ given in (\ref{alpha}) and
(\ref{beta}). Its estimation is also related to the minimum
nonzero eigenvalue of Laplacian $L_p$, which has been widely
studied in different situations (Merris, 1994).

In some existing works, including Jadbabaie, Lin, \& Morse (2003)
and Lin, Broucke, \& Francis  (2004), the convergence analysis
depends on theory of nonnegative matrices or stochastic matrices.
However, $F_{p}$ of system (\ref{model2}) fails to be transformed
easily to a matrix with some properties related to stochastic
matrices, and therefore, the effective methods used in Jadbabaie,
Lin, \& Morse (2003) or Lin, Broucke, \& Francis  (2004) may not
work. Here, we propose a Lyapunov-based approach to deal with the
problem.

\begin{thm}
\label{thm1} For any fixed $0<\gamma<1$ and $\bar\lambda$ defined
in (\ref{barlamb}), we take a constant
\begin{equation}
\label{constk} k> \frac{1}{4\gamma(1-\gamma^{2})\bar{\lambda}}.
\end{equation}
If the switching interconnection graph keeps connected, then
\begin{equation}
\label{result12} \lim_{t\to \infty}||\epsilon(t)||\leq C,
\end{equation}
for some constant $C$ depending on $\bar\delta$.  Moreover, if
$a(t)$ is known (i.e., $a(t)=a_0(t)$ or $\bar\delta= 0$),
\begin{equation}
\label{result11} \lim_{t\rightarrow \infty}\epsilon(t)=0.
\end{equation}
\end{thm}

Proof: Take a Lyapunov function
$V(\epsilon)=\epsilon^{T}(t)P\epsilon(t)$ with symmetric positive
definite matrix
\begin{equation}
\label{matrixp} P= \left(\begin{array}{cc}I_n&-\gamma I_n\\-\gamma
I_n&I_n\end{array}\right)\otimes I_2.
\end{equation}

The interconnection graph is time-varying, but the interconnection
graph associated with $F_p$ for some $p\in \mathcal{P}$ is
connected on an interval $[t_i, t_{i+1})$ with its topology
unchanged. Consider the derivative of $V(\epsilon)$:
\begin{equation}
\label{dotv}
\begin{split}
\dot{V}(\epsilon)|_{(\ref{model2})}&=\epsilon^{T}(F_{p}^{T}P+PF_{p})\epsilon+
2\epsilon^{T}F_pg \\
&\leq -\epsilon^{T}Q_{p}\epsilon+2(1+\gamma)\bar\delta
||\epsilon||
\end{split}
\end{equation}
where
\begin{equation}
\label{matrixq} Q_{p}=-(F_{p}^{T}P+PF_{p})=\begin{pmatrix}
2k(1-\gamma^{2})  M_{p}&-I_n\\
-I_n&2\gamma I_n
\end{pmatrix}\otimes I_2
\end{equation}
is a positive definite matrix because $2\gamma
I-\frac{1}{2k(1-\gamma^2)}M_{p}^{-1}$ and $M_p$ are positive
definite (by virtue of (\ref{constk}), Lemma \ref{lem1} and Lemma
\ref{lem3}).

Let $\mu_{i,j},i=1,\cdots,n, j=1,2$ denote the (at most) $2n$
different eigenvalues of $Q_p$ though $Q_p\in R^{4n\times 4n}$
defined in (\ref{matrixq}). Based on $\lambda_i(M_p)$, the
eigenvalues of $M_{p}$, we have the $2n$ eigenvalues in the
following forms:
\begin{equation*}
\begin{split}
&\mu_{i,1}=(1-\gamma^{2})k\lambda_i(M_p)+\gamma\\
&+\sqrt{[(1-\gamma^{2})k\lambda_i(M_p)+\gamma]^{2}-4\gamma(1-\gamma^{2})k\lambda_i(M_p)+1},
\end{split}
\end{equation*}
\begin{equation*}
\begin{split}
&\mu_{i,2}=(1-\gamma^{2})k\lambda_i(M_p)+\gamma\\
&-\sqrt{[(1-\gamma^{2})k\lambda_i(M_p)+\gamma]^{2}-4\gamma(1-\gamma^{2})k\lambda_i(M_p)+1},
\end{split}
\end{equation*}
for $i=1,...,n$.  Clearly, the smallest eigenvalue of $Q_p$ will
be found in the form of $\mu_{i,2}$ for some $i$.

Note that (\ref{constk}) implies $k\lambda_i(M_p)>
\frac{1}{4\gamma(1-\gamma^{2})}$. In this case, $\mu_{i,2}$
increases as $k\lambda_i(M_p)$ increases.  Therefore, the minimum
eigenvalue of $Q_p$ will be no less than
\begin{equation}
\label{barmu} \bar \mu=(1-\gamma^{2})k\bar \lambda+\gamma-
\sqrt{[(1-\gamma^{2})k\bar\lambda-\gamma]^{2}+1}>0,
\end{equation}
which is obtained by taking $\lambda_i(M_p)=\bar \lambda$ with a
given $k$ satisfying (\ref{constk}). In addition, since the
eigenvalues of $P$ are either $\mu_{min}=1-\gamma$ or
$\mu_{max}=1+\gamma$, we have
\begin{equation}
\label{relation} (1-\gamma)\|\e\|^2\leq V(\e) \leq
(1+\gamma)||\e||^2.
\end{equation}
Therefore,
$$
\min\frac{\epsilon^TQ_p\epsilon}{\epsilon^TP\epsilon}\geq
\frac{\bar\mu}{\mu_{max}} =2\beta,
$$
where $\beta=\frac{\bar \mu}{2(1+\gamma)}>0$ with $\bar\mu$
defined in (\ref{barmu}).

Due to (\ref{relation}),
$$
||\epsilon|| \leq \frac{1}{\sqrt{1-\gamma}}\sqrt{V(\epsilon)}.
$$
Therefore, from (\ref{dotv}),
\begin{equation*}
\begin{split} \dot V(\epsilon)|_{(\ref{model2})}&\leq -2\beta
V(\epsilon)+2\sqrt{\frac{(1+\gamma)^2V(\e)}{1-\gamma}}\bar\delta\\
&\leq -\beta
V(\epsilon)+\frac{(1+\gamma)^2\bar{\delta}^2}{(1-\gamma)\beta}
\end{split}
\end{equation*}
or equivalently,
\begin{equation*}
\begin{split}
 V(\epsilon(t))\leq
V(\epsilon(t_i))e^{-\beta(t-t_i)}+\frac{(1+\gamma)^2\bar\delta^2}{(1-\gamma)\beta^2}
&(1-e^{-\beta(t-t_i)}),\\
&t\in [t_{i}, t_{i+1}).
\end{split}
\end{equation*}
Thus, with $t_0=0$,
\begin{equation}
\label{expf} V(\epsilon(t))\leq V(\epsilon(0))e^{-\beta
t}+\frac{(1+\gamma)^2\bar\delta^2}{(1-\gamma)\beta^2}(1-e^{-\beta
t}),
\end{equation}
which implies (\ref{result12}) with taking
$C=\frac{1+\gamma}{(1-\gamma)\beta}\bar\delta$.

Furthermore, if $\bar\delta=0$, then (\ref{result11}) is obtained.
\hfill\rule{4pt}{8pt}

Next, we consider an extended case: the interconnection graph is
not always connected. Let $T>0$ be a (sufficient large) constant,
and then we have a sequence of interval $[T_j,T_{j+1}),\;
j=0,1,\cdots$ with $T_0=t_0,T_{j+1}=T_j+T$. Each interval
$[T_j,T_{j+1})$ consists of a number of intervals (still expressed
in the form of $[t_i,t_{i+1})$, during which the interconnection
graph is time-invariant), including the intervals during which the
graphs are connected and those during which the graphs are not. We
assume that there is a constant $\tau>0$, often called dwell time,
with $t_{i+1}-t_i\geq \tau, \;\forall i$.

Denote the total length of the intervals associated with the
connected graphs as $T_j^c$ in $[T_j,T_{j+1})$ and the total
length of the intervals with the unconnected graphs as $T_j^d$ in
$[T_j,T_{j+1})$. In what follows, we denote an upper bound of
$T_{j}^{d}\; (j=0,1,\cdots)$ as $T^d(<T)$, and a lower bound of
$T_j^c\; (j=0,1,\cdots)$ as $T^c(=T-T^d)$.

\begin{thm}
\label{thm2} During each time interval $[T_{j},T_{j+1})$, if the
total period that the interconnection graph is connected (i.e.,
$T^c$) is sufficient large, then (\ref{result12}) still holds with
$k$ given in (\ref{constk}). Moreover, (\ref{result11}) holds if
$\bar\delta= 0$ (or equivalently $a(t)=a_0(t)$).
\end{thm}

Proof: Still take a Lyapunov function
$V(\epsilon)=\epsilon^{T}P\epsilon$ with $P$ defined in
(\ref{matrixp}), and then we have (\ref{dotv}). If the graph
associated with $F_p$ for some $p\in \mathcal{P}$ is connected
during $[t_i,t_{i+1})$, then, according to Theorem \ref{thm1}, we
have
$$
V(\epsilon(t_{i+1}))\leq e^{-\beta(t_{i+1}-t_i)}
V(\epsilon(t_{i}))+ \frac{(1+\gamma)^{2}}{(1-\gamma)\beta^{2}}
\bar\delta^{2}.
$$

If the graph associated with $F_q$ for some $q\in \mathcal{P}$ is
not connected during $[t_l,t_{l+1})$. The minimum eigenvalue of
$Q_q$ is $\gamma-\sqrt{1+\gamma^{2}}(<0)$ and, by
(\ref{relation}), we have
$$
-\epsilon^{T} Q_q \epsilon \leq
(\sqrt{1+\gamma^2}-\gamma)\epsilon^{T} \epsilon \leq
\frac{\alpha}{2} V(\epsilon)
$$
where $\alpha=\frac{2\sqrt{1+\gamma^2}-2\gamma}{1-\gamma}$.

Similarly, with (\ref{relation}),
$$
\dot V(\epsilon(t))|_{(\ref{model2})}\leq \alpha V(\epsilon(t))+
\frac{2(1+\gamma)^{2}}{\alpha(1-\gamma)}\bar\delta^{2},\quad t\in
[t_l,t_{l+1}),
$$
and therefore,
\begin{equation}
V(\epsilon(t_{l+1}))\leq e^{\alpha (t_{l+1}-t_l)}
V(\epsilon(t_{l}))+
\frac{2(1+\gamma)^{2}}{\alpha^{2}(1-\gamma)}(e^{\alpha
T^d}-1)\bar\delta^{2}.
\end{equation}

Denote $\eta=\max\{\frac{(1+\gamma)^{2}}{\beta^{2}(1-\gamma)},
\frac{2(1+\gamma)^{2}}{\alpha^{2}(1-\gamma)}(e^{\alpha T^d}-1)\}$.
It is not hard to see that there are at most
$m_d=[\frac{T^d}{\tau}]+1$ intervals (in $[T_j,T_{j+1})$)
associated with unconnected graphs.  Therefore, we have
\begin{equation}
\begin{split} V(\epsilon(T_{j+1})) \leq& e^{-\beta
T_{j}^{c}+\alpha T_{j}^{d}
}V(\epsilon(T_{j}))+(1+e^{T^d}+e^{2T^d}\\
&+\cdots+e^{m_dT^d})\eta \bar\delta^2\\
 \leq &e^{-\beta T^{c}+\alpha
(T-T^{c}) }V(\epsilon(T_{j}))+\bar\eta\bar\delta^2
\end{split}
\end{equation}
with $\bar\eta=\frac{e^{(m_d+1)T^d}-1}{e^{T^d}-1}\eta>0$.

If $\beta T_{c}>\alpha (T-T^{c})$ or $T_c>\frac{\alpha
T}{\alpha+\beta}$, then $\nu=e^{-\beta T^{c}+\alpha (T-T^{c})}<1$.
Thus,
\begin{align*}
V(\epsilon(T_{j+1})) &\leq
\nu^{j+1}V(\epsilon(T_{0}))+(\nu^{j}+\cdots+1)\bar\eta\bar\delta^2\\
&\leq
\nu^{j+1}V(\epsilon(T_{0}))+\frac{1-\nu^{j+1}}{1-\nu}\bar\eta\bar\delta^2.
\end{align*}
For any $t>0$, there is $j$ such that $T_j<t<T_{j+1}$ with
$$
V(\epsilon(t)) \leq e^{\alpha T^{d}}
V(\epsilon(T_{j}))+\bar\eta\bar\delta^2
$$
Thus, (\ref{result12}) is obtained with taking
$C=\sqrt{\frac{(e^{\alpha T^{d}}+1-\nu)
\bar\eta}{(1-\nu)(1-\gamma)}}\bar\delta$.

Furthermore, if $\bar\delta=0$, then $C=0$, which implies
(\ref{result11}), or $\epsilon \to 0$ as $t\to \infty$.
\hfill\rule{4pt}{8pt}

In fact, the proposed estimation idea can be extended to the case
of an active leader with the following dynamics:
\begin{equation}
\label{general}
\begin{cases}
\dot{x}_0^1=x_0^2 \\
\dot{x}_0^2=x_0^3\\
\vdots\\
\dot{x}_0^{\kappa}=a(t)=a_0(t)+\delta(t)\\
y=x_0=x_0^1\in R^2
\end{cases}
\end{equation}
where $y(t)$ is the measured output variable of the leader and
$a(t)$ is its input variable.  The dynamics of each agent is still
taken in the form of (\ref{modela}).  Then we will construct an
observer as we did for system (\ref{model0}). Here, for the space
limitations, we only give the corresponding error system, which
can be expressed as: {\scriptsize
$$
\begin{pmatrix} \dot{\bar x}^1\\
\dot{\bar x}^2\\\vdots\\ \dot{\bar x}^{\kappa}\end{pmatrix}=
\begin{pmatrix}-kM_{p}&I_n&&\\
-\gamma_1 kM_{p}&0&I_n&\\
\vdots &  &&\\
-\gamma_{\kappa-2}k M_{p}&&0&I_n\\
-\gamma_{\kappa-1}k M_{p}&&&0
\end{pmatrix}\otimes I_2\begin{pmatrix} {\bar x^1}\\
{\bar x}^2\\\vdots\\ {\bar x}^{\kappa}\end{pmatrix}+\begin{pmatrix} 0\\
0\\\vdots\\ - {\bf 1}\otimes \delta\end{pmatrix}
$$ }
or equivalently in a compact form:
$$
\dot\epsilon=F_p\epsilon+g\in R^{2n\kappa}
$$
where $k>0$ and $0<\gamma_j<1, \; (j=1,...,\kappa-1)$ are suitable
real numbers and ${\bar x}^i=x^i-\textbf{1}\otimes x_{0}^i\in
R^{2n}$ with $x^1=x$ and $x^i\; (2\leq i\leq \kappa)$ as the
vector whose components are the respective estimated values of
$x_0^i$ by $n$ agents.

To obtain the results similar to Theorems \ref{thm1} or
\ref{thm2}, we need to find a suitable quadratic Lyapunov
function; that is, to construct an appropriate positive definite
matrix $P$ such that $F_{p}^{T}P+PF_{p}$ is negative definite then
the corresponding graph $\bar {\mathcal G}_p$ is connected. For
example, when $\kappa=3$, we can choose {\scriptsize
\begin{equation*}
F_p=\left(\begin{array}{ccc}
-kM_{p}&I_n&0\\
-\frac{8k}{9}M_{p}&0&I_n\\
-\frac{4k}{9}M_{p}&0&0
\end{array}\right)\otimes I_2,\; P=\left(\begin{array}{ccc}I_n&-\frac{2}{3}I&0
\\-\frac{2}{3}I_n&I_n&-\frac{1}{2}I_n\\0&-\frac{1}{2}I_n&I_n\\\end{array}\right)\otimes
I_2.
\end{equation*}}
%When $n=4$, we can choose??
%$$F_p=\left(\begin{array}{cccc} M_{p}&I_n&0&0\\
%\frac{48}{35}M_{p}&0&I_n&0\\\frac{9}{7}M_{p}&0&0&I_n\\
%\frac{27}{35}M_{p}&0&0&0\end{array}\right)\otimes I_2,
%P=\left(\begin{array}{cccc}I_n&-\frac{3}{5}I_n&0&0
%\\-\frac{3}{5}I_n&I_n&-\frac{3}{5}I_n&0\\0&-\frac{3}{5}I_n&I_n&-\frac{3}{5}I_n\\
%0&0&-\frac{3}{5}I_n&I_n\end{array}\right)\otimes I_2.$$

\section{Conclusions}
This paper studied the consensus problem of a group of autonomous
agents with an active leader, whose velocity cannot be measured.
To solve the problem, a distributed feedback (i.e., (\ref{con}))
along with a distributed state-estimation rule (i.e.,
(\ref{update1})) was proposed for each continuous-time dynamical
agent in a varying interconnection topology. In fact, generalized
cases, including those when the interconnection graphs are
directed and when the dynamics of each agent is more complex than
system (\ref{modela}), will be considered in future.

%\section*{Acknowledgement}
\begin{ack}
The authors wish to thank the reviewers for their constructive
suggestions, especially the reviewer who introduced us a framework
of distributed observer design for multi-agent systems. This work
was supported by the NNSF of China under Grants 60425307, 50595411
and 60221301.
\end{ack}
\section*{References}
%\bibliographystyle{plain}
%\bibliography{agent_hong}
%\begin{thebibliography}{99}
\begin{description}
{\small

 \item Fax, A., \& Murray, R. M. (2004). Information flow
and cooperative control of vehicle formations, {\it IEEE
Transactions on Automatic Control}, {\it 49}(9), 1453-1464.

\item Godsil, C., \& Royle, G., (2001). {\it Algebraic Graph
Theory,} New York: Springer-Verlag.

\item Horn, R., \& Johnson, C. (1985). {\it Matrix Analysis,} New
York: Cambbridge Univ. Press.

\item Jadbabaie, A.,  Lin, J., \& Morse, A. S. (2003).
Coordination of groups of mobile autonomous agents using nearest
neighbor rules, {\it IEEE Trans. Automatic Control}, {\it 48}(6),
998-1001.

\item Merris, R. (1994). Laplacian matrices of graphs: a survey,
{\it Linear Algebra and Applications}, {\it 197}, 143-176.

\item Lin, Z., Broucke, M., \& Francis, B. (2004). Local control
strategies for groups of mobile autonomous agents, {\it IEEE
Trans. Automatic Control}, {\it 49}(4), 622-629.

%\bibitem \"{O}gren, P., Egerstedt, E., \& Hu, X. (2002). A control
%Lyapunov approach to multiagent coordination, {\it IEEE Trans.
%Robotics and Automation}, {\it 18}(5), 847-851.

\item Olfati-Saber, R., \& Murray, R. M. (2004). Consensus
problems in networks of agents with switching topology and
time-delays, {\it IEEE Trans. Automatic Control}, {\it 49}(9),
1520-1533.

%\bibitem{line} Kailath, T. {\it Linear Systems},
%Englewood Cliffs: Prentice-Hall, Inc., 1980.

\item Savkin, A. (2004). Coordinated collective motion of groups
of autonmous mobile robots: analysis of Vicsek's model, {\it IEEE
Trans. Automatic Control}, {\it 49}(6), 981-983.}
\end{description}
%\end{thebibliography}
\end{document}